\documentclass[sort&compress]{aipproc}
\layoutstyle{8x11double}

\begin{document}
\title{Vector Boson Scattering at High Mass with ATLAS}
\author{Adam Davison}{
	address={on behalf of the ATLAS Collaboration}
	,altaddress={Department of Physics \& Astronomy, University College London}
}

\begin{abstract}
	In the absence of a light Higgs boson, the mechanism of electroweak symmetry breaking will be best studied in processes of vector boson scattering at high mass. Various models predict resonances in this channel. Scalar and vector resonances have been investigated in the $WW$, $WZ$ and $ZZ$ channels. The ability of ATLAS to measure the di-boson cross-section over a range of centre-of-mass energies has been studied with particular attention paid to the reconstruction of jet pairs with low opening angle resulting from the decays of highly boosted vector bosons.
\end{abstract}

\maketitle

In the absence of a light Higgs boson, the Standard Model predicts unphysical cross-sections for vector boson scattering at the TeV scale~\cite{Dawson:1998yi}. Some mechanism for Electroweak Symmetry Breaking must be observed at the LHC to resolve this inconsistency. The Standard Model Higgs is just one of many possible solutions to this problem~\cite{Lane:2002sm}\cite{Csaki:2003dt,Csaki:2003zu,Cacciapaglia:2004rb}\cite{Csaki:2004sz,Birkedal:2005yg,Sekhar_Chivukula:2006cg}\cite{Casalbuoni:2000gn} and measurements of vector boson scattering cross-sections will likely be a key tool in distinguishing between different models. It is therefore desirable to understand the main issues with performing such an analysis with ATLAS and also have some feel for the required luminosity for such measurements to be feasible.

\section{Experimental Signature}
The category of processes described here as vector boson scattering includes a range of possible final states. We are scattering a pair of vector bosons, which can be $WW$, $WZ$ or $ZZ$. Furthermore, each boson may decay leptonically or hadronically. Obviously some channels are more challenging than others. Usually channels with all-hadronic decays are inaccessible due to large backgrounds from QCD multijet processes. Channels with both bosons decaying leptonically tend to be extremely clean but require higher luminosities due to low branching ratios. Our attention then turns to channels where one boson decays hadronically and one leptonically. These provide a challenging but promising arena in which to study vector boson scattering and also an opportunity to take advantage of ATLAS' comparatively fine hadronic calorimetry.

The main backgrounds to observation of these processes are $W/Z+jets$ and $t\overline{t}$ production. In $W/Z+jets$ the jets may simulate the presence of a hadronic vector boson. $t\overline{t}$ events are an issue for $WW$ channels due to the presence of two real $W$ bosons, although the signal has a significantly different event topology. The presence of these backgrounds means these channels are hard to observe in the low mass region. As a general guide, the $p_\perp$ of the vector bosons is usually required to be above $200$ GeV, which presents unique challenges for identifying the hadronic system, as will be discussed in more detail below.

The experimental signature of semi-leptonic vector boson scattering events is a leptonic system reconstructible as either a $W$ or $Z$, a hadronic system from the hadronic decay and two "tag" jets from the original boson production. The emission of high-$p_\perp$ vector bosons from the incoming partons tends to be balanced by the presence of jets at very low angle. While other processes may produce similar signatures in the center of the detector they are not correlated with the presence of these "tag" jets, making this a powerful additional discriminator.

\section{Hadronic Vector Boson Identification}
For the first time at the LHC, we will have an opportunity to observe the production of $O(100)$ GeV particles with significant boosts. Decay products in this scenario tend to be much closer together which presents a challenge in terms of identifying separate objects in a detector. However, combinatoric effects are reduced or even eliminated in these scenarios. Specifically for hadronic decays, particles start to be resolved as a single jet which means that standard techniques such as dijet mass cuts are no longer applicable.

\begin{figure}
	\centering
	\includegraphics[width=\linewidth]{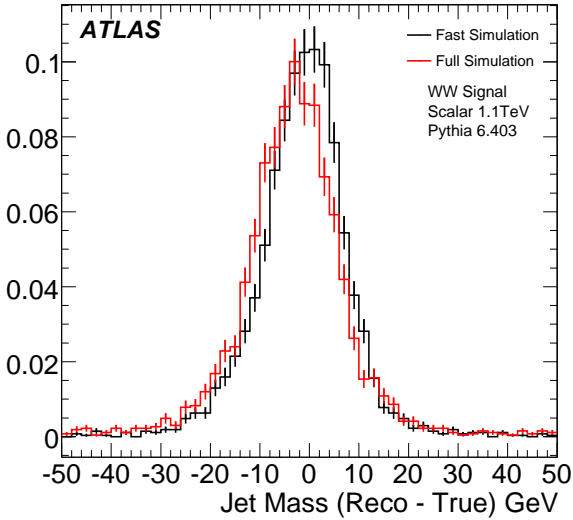}
	\caption{Resolution of single jet mass in a sample of boosted $W$ bosons from a vector boson scattering signal as observed in ATLAS simulation~\cite{CSC}.}
	\label{fig:mres}
\end{figure}

The first technique at our disposal is single jet mass. By analogy with traditional dijet mass cuts, if the decay products are all boosted into a single jet, then the mass of that jet should be of the order of the mass of the parent particle. By contrast, jet mass in QCD processes tends to be much lower. This provides a powerful tool for identifying jets containing highly boosted heavy particle decays. The ability of the ATLAS detector to measure these quantities is explored through plots such as Figure \ref{fig:mres}, showing a resolution of $O(10)$ GeV, enough to possibly even separate $W$ and $Z$ bosons with high enough statistics.

However, it is also useful to have some additional discriminating power. Specifically, when using the $k_\perp$ jet finder, it is possible to explore the energy scales at which the jet can be decomposed into smaller sub-jets~\cite{Butterworth:2002tt}. These "y-scales" are an additional useful discriminator because while QCD processes can produce high mass jets, often the source is multiple smaller showering splittings, whereas with the decay of a heavy particle there must be a hard splitting. Therefore we can examine the scale at which a jet can be subdivided into two sub-jets and for vector bosons we find it to be $O(m_{W/Z})$ while for QCD jets it tends to be much lower. These "y-scales" are highly correlated with the jet mass but some significant additional rejection is still provided by a cut on this value.

These techniques enable heavy jets which are likely to have come from vector boson decay to be effectively selected. Recent advances in jet reconstruction propose that other related methods may be of value for this channel but this has yet to be studied in this context.

\section{ATLAS Simulation}
This process has now been studied for the first time with full detector simulation of the ATLAS detector~\cite{CSC}. Monte Carlo samples of signals and backgrounds have been simulated and a sample analysis performed to explore the sensitivity of the ATLAS experiment.

\begin{figure}
	\centering
	\includegraphics[width=\linewidth]{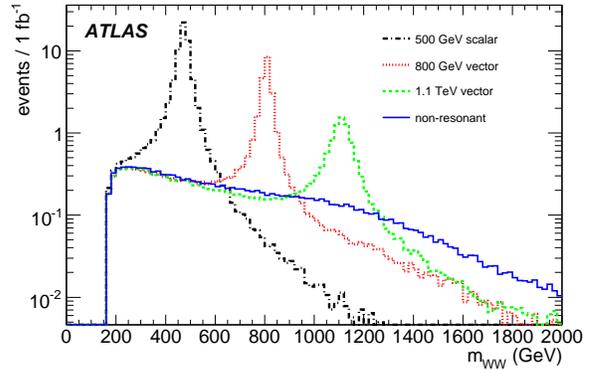}
	\caption{Parton level differential cross-sections for a selection of possible vector boson scattering signal samples.}
	\label{fig:signals}
\end{figure}

Due to the broad range of theoretical scenarios of interest in this channel, this study attempted to produce model independent results which can then be easily interpreted in many contexts. Specifically, the analysis is designed to produce invariant mass spectra of the vector boson scattering system, where resonances or other more generic predictions can be compared to data, like those in Figure \ref{fig:signals}.

Signals were generated by modifying couplings in the EWChL model~\cite{Kilian:2003pc,Kilian:2003yw} and applying the Padé unitarization protocol~\cite{Dobado:1996ps}. Depending on the choice of coupling constants, this produces scalar or vector resonances. This was implemented using the {\sc Pythia}~\cite{pythia} generator. Backgrounds of $W/Z+jets$ and $t\overline{t}$ events were produced with MadGraph~\cite{Maltoni:2002qb} and MC@NLO~\cite{mcatnlo2} respectively.

\section{Sensitivity}
The sensitivity to a selection of vector and scalar resonances, with masses in the range $500$ GeV to $1.1$ TeV was explored. The requirements of the analysis were broadly the following:

\begin{itemize}
\item 1 hadronic $W$ or $Z$ with $p_\perp > 200$ GeV and $|\eta| < 2$,
\item 1 leptonic $W$ or $Z$ with $p_\perp > 200$ GeV and $|\eta| < 2$,
\item 2 "tag" jets present in opposite hemispheres of the detector with $|\eta| > 2$,
\item No top candidates,
\item No additional jets with $|\eta| < 2$,
\end{itemize}

where the hadronic vector boson has been identified as either a single jet with high mass (and "y-scale" when the $k_\perp$ algorithm is used) as described above or as a pair of jets with high dijet mass.

Using this selection over several channels, the sensitivities observed in Table \ref{table:sens} were extracted for observation of peaks due to resonances. The fully leptonic modes perform better for lower mass, higher cross-section resonances whereas the semi-leptonic modes perform better for the higher mass resonances. These results imply that ATLAS is sensitive to some types of resonances with as little as a few tens of fb$^{-1}$ of well understood data. However models which predict multiple resonances or resonance production with higher cross-sections or in multiple channels could be observed sooner.

\begin{table}
\begin{tabular}{ccc}
\hline
\textbf{Channel} & \textbf{Resonance Mass} & \parbox[][][c]{0.7in}{\centering \textbf{Lumi for $3\sigma$ evidence}} \\ \hline
$WW/WZ\rightarrow l\nu jj$ & $500$ GeV & $85fb^{-1}$ \\ \hline
$WW/WZ\rightarrow l\nu jj$ & $800$ GeV & $20fb^{-1}$ \\ \hline
$WW/WZ\rightarrow l\nu jj$ & $1.1$ TeV & $85fb^{-1}$ \\ \hline
$ZW/ZZ\rightarrow ll jj$ & $500$ GeV & $30fb^{-1}$ \\ \hline
$ZW/ZZ\rightarrow ll jj$ & $800$ GeV & $30fb^{-1}$ \\ \hline
$ZW/ZZ\rightarrow ll jj$ & $1.1$ TeV & $90fb^{-1}$ \\ \hline
$ZZ\rightarrow \nu \nu ll$ & $500$ GeV & $20fb^{-1}$ \\
\hline
\end{tabular}
\caption{Estimated sensitivities of the ATLAS experiment for various resonances in different channels.}
\label{table:sens}
\end{table}

\section{Conclusions}
Vector boson scattering has been presented as an important channel for study at the LHC. Although in many ways challenging, especially in terms of requirements in hadronic calorimetry, it may be the key to understanding particle physics at the TeV scale.

This channel has now been studied in some detail using the ATLAS detector simulation and has been found to have potential for discovery within a few years of design luminosity delivered to ATLAS from the LHC.

\bibliographystyle{aipproc}
\bibliography{VBS-CSC-note}

\end{document}